\documentclass[12pt]{article}
\usepackage{epsfig,amssymb}
\begin{document}

\begin{flushright}
{\tt hep-th/0404163}
\end{flushright}

\vspace{5mm}

\begin{center}
{{{\Large \bf Tubular D-branes in Salam-Sezgin Model}}\\[14mm]
{Chanju Kim}\\[3mm]
{\it Department of Physics, Ewha Womans University,
Seoul 120-750, Korea}\\
{\tt cjkim@ewha.ac.kr}\\[7mm]
{Yoonbai Kim \ and \ O-Kab Kwon}\\[3mm]
{\it BK21 Physics Research Division and Institute of
Basic Science,\\
Sungkyunkwan University, Suwon 440-746, Korea}\\
{\tt yoonbai@skku.edu~~okab@skku.edu}
}
\end{center}
\vspace{10mm}

\begin{abstract}
We study DBI-type effective theory of an unstable D3-brane in the background
manifold ${\rm R}^{1,1}\times {\cal M}_2$ where ${\cal M}_2$ is an arbitrary
two-dimensional manifold.
We obtain an exact tubular D2-brane solution of arbitrary cross sectional
shape by employing $1/\cosh$ tachyon potential. 
When ${\cal M}_2={\rm S}^{2}$, the solution is embedded in the background 
geometry ${\rm R^{1,3}} \times {\rm S}^2$ of Salam-Sezgin model.
This tachyon potential
shows a unique property that an array of tachyon soliton
solutions has a fixed period which is independent of integration constants of
the equations of motion.
The thin BPS limit of the configurations
leads to supertubes of arbitrary cross sectional shapes.
\end{abstract}

\newpage

\setcounter{equation}{0}
\section{Introduction}

When D-branes are wrapped on some nonsupersymmetric cycles in the moduli
space of compactified manifolds, e.g., K3 or Calabi-Yau manifolds,
they are dissociated and form several stable D-branes among which each
is wrapped on a supersymmetric cycle~\cite{Sen:1998ex,Horava:1998jy}.
A representative example is a D$p$-brane and one of the directions is
compactified as a circle of radius $R$.
At the critical radius $R=\sqrt{2}$, it decays into a pair of
D$(p-1)\bar{{\rm D}}(p-1)$ branes~\cite{Sen:1998tt}
situated at diametrically opposite
points. In terms of boundary conformal field theory (BCFT),
this phenomenon is described by a marginal deformation interpolating
between the original unstable D$p$-brane and the
D$(p-1)\bar{{\rm D}}(p-1)$ pair~\cite{Sen:1998ex}.

In the context of effective field theory (EFT) where the instability of
D$p$-brane is represented by condensation of a tachyon
field, the Dirac-Born-Infeld (DBI) type
effective action~\cite{Sen:1999md,Garousi:2000tr}
with the choice of $1/\cosh$
potential~\cite{Buchel:2002tj,Kim:2003he,Leblond:2003db}
can reproduce this as an array of static tachyon kink-antikinks with fixed
period $2\pi R$~\cite{Lambert:2003zr,Kim:2003in}. Only in this EFT,
the tachyon profile in BCFT and the tachyon field configuration in EFT has
one-to-one correspondence by an explicit point
transformation~\cite{Kutasov:2003er} so that the identification between
them can clearly be made at the classical level~\cite{Sen:2003bc}. 
When the gauge field is turned on on the unstable D-brane, 
the period of the D$(p-1)\bar{{\rm D}}(p-1)$ array starts changing 
from $2\pi R$ to a larger value and it eventually goes to infinity when
the electric field approaches the critical 
value~\cite{Kim:2003in,Sen:2003bc,Kim:2003ma}. 
For each value of the electric field,
the period is fixed and independent of other integration constants of
the equation of motion in EFT with $1/\cosh$
potential~\cite{Kim:2003in,Kim:2003ma}.

On the other hand, when the shape of tachyon potential is chosen to
be different from $1/\cosh$,
the property of fixed periodicity seems likely to be
lost even in pure tachyon case~\cite{Brax:2003rs}.
In this paper, we will show that, in the EFT with DBI
type action, the $1/\cosh$ tachyon potential should uniquely be chosen
in order to keep this periodic property.

The codimension-one D-branes which are represented as kinks or
antikinks in EFT become BPS objects only in zero 
thickness limit~\cite{Sen:2003tm} (except the case of the composite 
of D$p$-brane and fundamental string 
(F1) fluid with critical electric field~\cite{Kim:2003in,Kim:2003ma}).
This is also true for codimension two or three objects such as
vortex-antivortex pairs or monopole-antimonopole pairs 
though BCFT description may not be
explicit~\cite{Sen:1998ex,Horava:1998jy}.

In this paper we will consider another configuration, a generation of
a tubular brane on R${}^{1}\times$S${}^{2}$ of which the thin limit
is a supertube~\cite{Mateos:2001qs} along the equator of S${}^{2}$.
The tubular D2-brane can have an arbitrary cross sectional
shape~\cite{Bak:2001xx} and it is natural to expect this property to appear 
also in tachyon tubes from an unstable D3-brane~\cite{Kim:2003uc}. 
We will find thick tachyon tube solutions of arbitrary cross section from an 
unstable D3-brane on the background manifold R${}^{1,1}\times {\cal M}_{2}$,
and discuss BPS supertube limit by taking the thickness to be zero.

The case of ${\cal M}_{2}={\rm S}^{2}$ is of our particular interest.
In this case the base spacetime is embedded in
R${}^{1,3}\times$S${}^{2}$ of six-dimensional Salam-Sezgin
model~\cite{Salam:1984cj}. 
Recently Salam-Sezgin model has been studied in various
contexts~\cite{Aghababaie:2002be}--\cite{Cvetic:2003xr}.
In particular, several investigations have been made in relation with 
the vacuum structure.
A consistent S${}^{2}$ reduction of the Salam-Sezgin model was performed
and its four-dimensional spectrum was analyzed~\cite{Gibbons:2003gp}.
A new family of supersymmetric vacua in the six-dimensional chiral 
gauged $N=(1,0)$ supergravity was discovered, of which the generic form
is AdS${}_{3}\times$S${}^{3}$, and in this scheme R${}^{1,3}\times$S${}^{2}$
can be viewed as a fine-tuning~\cite{Guven:2003uw}.
Uniqueness of the Salam-Sezgin vacuum among all nonsingular
backgrounds with four-dimensional Poincar\'{e}, de Sitter, or anti de
Sitter invariance was proved~\cite{Gibbons:2003di}.

Our analysis is based on the EFT, and  it is unclear whether or not 
the obtained tube solution can be a consistent BCFT solution in the 
background of string theory. Since 
higher-dimensional origin of Salam-Sezgin model has also been 
obtained~\cite{Kerimo:2003am,Cvetic:2003xr}, this important issue should be
addressed in a consistent manner.

The rest of paper is organized as follows. In section 2, we prove uniqueness
of the tachyon potential for fixed periodicity of the array of tachyon
soliton-antisoliton pairs. 
In section 3, we obtain exact
tachyon tube solutions on ${\rm R}^{1,1}\times {\cal M}_2$, where
${\cal M}_2$ is a two-dimensional manifold, and discuss their BPS limit. 
In section 4, we consider the case ${\cal M}_2 = {\rm S}^2$ in more detail.
We conclude in section 5 with brief discussion.

\setcounter{equation}{0}
\section{Tachyon Potential of D-brane Wrapped on a Cycle}

The effective tachyon action for an unstable D$p$-brane
system~\cite{Sen:1999md,Garousi:2000tr} is
\begin{equation}\label{fa}
S= -{\cal T}_{p} \int d^{p+1}x\; V(T) \sqrt{-\det (g_{\mu\nu} + F_{\mu\nu} +
\partial_\mu T\partial_\nu T )}\, ,
\end{equation}
where $g_{\mu\nu}$ is the metric given from the closed string sector,
$T(x)$ is tachyon field, and $F_{\mu\nu}$ field strength tensor of a gauge
field $A_{\mu}$ on the D$p$-brane, of which the constant piece can also be
interpreted as NS-NS two form field.
We set $2\pi \alpha'=1$ and then ${\cal T}_{p}$ is tension of the D$p$-brane.

Since tachyon potential measures variable tension of the unstable D-brane,
it should be a runaway potential connecting
\begin{equation}\label{vbd}
V(T=0)=1~~\mbox{and}~~ V(T=\infty)=0.
\end{equation}
Various forms of it have been proposed, e.g., $V(T)\sim e^{-T^{2}}$ from
boundary string field theory~\cite{Gerasimov:2000zp} or $V(T)\sim e^{-T}$ for
large $T$ in Ref.~\cite{Sen:2002an}. In this paper, we employ the
form~\cite{Buchel:2002tj,Kim:2003he,Leblond:2003db}
\begin{equation}\label{V3}
V(T)=\frac{1}{\cosh \left(\frac{T}{R}\right)}
\end{equation}
which connects the small and the large $T$ behaviors smoothly.
Here, $R$ is $\sqrt{2}$ for the non-BPS D-brane in the superstring
and 2 for the bosonic string.
This form of the potential has been derived in open string theory
by taking into account
the fluctuations around $\frac{1}{2}$S-brane configuration
with the higher derivatives neglected, i.e., $\partial^2 T = \partial^3 T=
\cdots = 0$~\cite{Kutasov:2003er,Okuyama:2003wm,Niarchos:2004rw}.

Most of the physics of tachyon condensation is irrelevant to the detailed
form of the potential once it satisfies the runaway property and the
boundary values (\ref{vbd}). For example,
both the basic runaway behavior of rolling tachyon solutions~\cite{Sen:2002an}
and the BPS nature of tachyon kinks with zero thickness~\cite{Sen:2003tm}
are attained irrespective of the specific shape of the potential which
just reflects a detailed decaying dynamics of the unstable D-brane.

On the other hand, there are also some nice features of the form
(\ref{V3}) in addition to the fact that it is derived from open string
theory in a specific regime.
Under the 1/cosh tachyon potential (\ref{V3}),
exact solutions are obtained for rolling
tachyon~\cite{Kim:2003he,Kim:2003ma} and tachyon kink
solutions on unstable D$p$ with a coupling of abelian gauge field
for arbitrary $p$~\cite{Lambert:2003zr,Kim:2003in,Brax:2003rs,
Kim:2003ma,Kim:2003uc}.
Another useful property may be the observation
that some of the obtained classical solutions $T(x)$ in the EFT
(\ref{fa}), e.g., rolling tachyons~\cite{Lambert:2003zr} and 
tachyon kinks~\cite{Sen:2003bc}, can be directly translated to BCFT 
tachyon profiles $\tau(x)$ in open string
theory described by the following relation obtained in 
Ref.~\cite{Kutasov:2003er}, 
\begin{equation}\label{oto}
\frac{\tau(x)}{R}=\sinh \left(\frac{T(x)}{R}\right).
\end{equation}

In this section, we would like to discuss another important feature
of the 1/cosh potential (\ref{V3}), which is not shared by any other form.
Among the tachyon soliton solutions in the effective theory, 
various tachyon array solutions of codimension one have been
found, namely, those formed by pure tachyon kink-antikink~\cite{Sen:1998tt,
Sen:1998ex,Lambert:2003zr,Kim:2003in}, tachyon kink-antikink
coupled to the electromagnetic field~\cite{Kim:2003in,Sen:2003bc,Kim:2003ma},
and tachyon tube-antitube~\cite{Kim:2003uc}.
An interesting property of all these solutions is that,
with the 1/cosh potential in the EFT,
the periodicity of the array is independent of any integration constant
of the equation of motion, much like the case of simple harmonic oscillator.
Here we will show that the converse is also true by adopting the similar
line of argument
to the case of simple harmonic oscillator: imposing the condition that the
periodicity of the tachyon array solutions should be independent of 
the integration constant of the equation of motion uniquely
determines the tachyon potential as Eq.~(\ref{V3}). This property is
necessary if one wishes to identify the array solution as a configuration on
a circle or a sphere of a fixed radius~\cite{Sen:2003bc,Sen:2003zf}.

To begin with, we recall that the relevant equation for all the array
solutions with $T=T(x)$ and $F_{\mu\nu}$ is summarized by a
single first-order ordinary differential equation
\begin{equation}\label{eee}
{\cal E} = \frac12 T'^2 + \frac{1}{h} U(T),
\end{equation}
where $U = V^{2}(T)$. (See Ref.~\cite{Kim:2003in,Kim:2003ma,Kim:2003uc}
and also (\ref{trr}) in the next section.)
For the array of kink-antikink, two parameters ${\cal E}$ and $h$ are
\begin{equation} \label{uv}
{\cal E} = -\frac{\beta_p}{2\alpha_p},\qquad
h = -\frac{2\alpha_p \gamma_p^2}{{\cal T}_p^2},
\end{equation}
where $\beta_p=-\det(\eta_{\mu\nu}+F_{\mu\nu})$, $\alpha_{p}$ is cofactor of
11-component of the matrix $-(\eta+F)_{\mu\nu}$, and
$\gamma_{p}$ an integration constant~\cite{Kim:2003in,Kim:2003ma}.
For the array of tube-antitube, ${\cal E}$ and $h$ are
\begin{equation} \label{uv1}
{\cal E} = -\frac{1}{2},\qquad
h = -\frac{2\alpha^2\beta^2}{{\cal T}_3^2},
\end{equation}
where $\alpha$ is D0 charge density per unit length and $\beta$ an
integration constant~\cite{Kim:2003uc}. (See also (\ref{trr})
in the next section.)
Then, for our purpose, the coefficient $h$ in front of the potential is
negative and to be varied, and ${\cal E}$ is regarded as a constant
with $1/h< {\cal E}<0$.

Let us require the period to be independent of $h$ in Eq.~(\ref{eee}).
Denoting the period as $\zeta$, we have
\begin{eqnarray}
\frac{\pi}{2} \zeta
&=& \int_0^{T_{\rm max}} \frac{dT}{\sqrt{2[{\cal E} - U(T)/h]}} \nonumber \\
&=& \int_{U_0}^{h{\cal E}} \frac{dT/dU}{\sqrt{2({\cal E} - U/h)}}dU,
\label{peq}
\end{eqnarray}
where $T_{\rm max}$ is the maximum value of the tachyon field,
$U(T_{\rm max}) = h {\cal E}$, and $U_0 = U(T=0)$.
It turns out to be convenient to define the variable $\eta = h{\cal E}$.
Then Eq.~(\ref{peq}) becomes
\begin{equation}
\frac{\pi}{2} \zeta
= \frac{1}{\sqrt{2|{\cal E}|}} \int_{U_0}^\eta
  \frac{\sqrt{-\eta}}{\sqrt{\eta - U}} \frac{dT}{dU} dU.
\end{equation}
If both sides of this equation are divided by $\sqrt{(-\eta)(U-\eta)}$ and
integrated with respect to $\eta$ from $U_0$ to $U$,
\begin{eqnarray}
\frac{\pi \zeta}{2} \int_{U_0}^U \frac{d\eta}{\sqrt{\eta^2 - U \eta}}
&=& \frac{1}{\sqrt{2|{\cal E}|}} \int_{U_0}^U  \int_{U_0}^\eta d\eta dU'
  \frac{dT(U')/dU'}{\sqrt{(U-\eta)(\eta - U')}} \nonumber \\
&=& \frac{1}{\sqrt{2|{\cal E}|}} \int_{U_0}^U dU' \int_{U'}^U
  d\eta \frac{dT(U')/dU'}{\sqrt{(U-\eta)(\eta - U')}},
\label{pint}
\end{eqnarray}
where we changed the order of integration in the second line.

It is now elementary to perform the integral (\ref{pint}) in the both sides.
The result is
\begin{equation}\label{Uf}
\pi\zeta \, {\rm arccosh}\left( \sqrt{\frac{U_0}{U}} \right)
= \frac{\pi T}{\sqrt{2|{\cal E}|}},
\end{equation}
i.e.,
\begin{equation} \label{UT}
U(T) = \frac{U_0}{\cosh^2(T/R)},
\end{equation}
where $R = \zeta \sqrt{2|{\cal E}|}$. Comparing Eq.~(\ref{Uf}) with
Eq.~(\ref{eee}), we see that $V(T) = 1/\cosh(T/R)$ as asserted.
This property can also be seen clearly after a point transformation
(\ref{oto}) to the equation (\ref{eee}), which, under the specific
tachyon potential (\ref{V3}), results in
\begin{equation}\label{eef}
{\cal E}'=\frac{1}{2}\tau'^{2}+\frac{1}{2}\omega^{2}\tau^{2},
\end{equation}
where $0<{\cal E}'=-1/h+{\cal E}$ and $0<\omega^{2}=-2{\cal E}/R^{2}$.
Since both ${\cal E}$ and $R$ are fixed but $h$ is a variable, 
${\cal E}'$ is a positive variable and $\omega^{2}$ a constant. Therefore, 
Eq.~(\ref{eef}) is formally equivalent to the expression of the
mechanical energy ${\cal E}'$ of a 1-dimensional simple harmonic 
oscillator with unit mass of which the position is $\tau$ at time $x$. 
According to the proof in Ref.~\cite{LLt}, its period $2\pi/\omega$
is independent of the value of ${\cal E}'$ only for the
simple harmonic oscillator.

This periodic property of the array configurations in the effective field
theory is desirable
if we wish to identify the array solution as a pair of
D$(p-1)\bar{{\rm D}}(p-1)$ obtained from an unstable
D$p$-brane wrapped on a cycle in the context of string
theory~\cite{Sen:1998ex,Horava:1998jy,Sen:2003bc}.
(Note also from Eq.~(\ref{UT}) that the compactified length $\zeta$ varies
as the electromagnetic field changes.)
In this sense, our proof in this section tells the uniqueness of the 1/cosh
tachyon potential (\ref{V3}) for the tachyon field in Eq.~(\ref{fa})
in studying the generation of
codimension one extended objects on nonsupersymmetric cycles.
In section 3, we will find a family of tachyon tube solutions with such
periodicity on
R${}^{1}\times {\cal M}_{2}$, and
in section 4, will demonstrate that single
tachyon tube on ${\rm R}^{1}\times {\rm S}^{2}$
forms a thin tubular object of which the
geometry is ${\rm R}^{1}\times {\rm S}^{1}$ in the BPS limit.

\setcounter{equation}{0}
\section{Tachyon Tubes of Arbitrary Cross Section and BPS Limit}

In this section we consider tachyon tube configurations in the theory
described by
the DBI type action (\ref{fa}) on ${\rm R}^{1,1} \times {\cal M}_2$ in
the coordinate system $(t,z,u,v)$ with metric
\begin{equation}\label{met}
ds^2 = -dt^2 + dz^2 + du^2 +f(u)^{2}dv^{2},
\end{equation}
where $f(u)$ is an arbitrary function. Depending on $f(u)$
the two-dimensional manifold ${\cal M}_{2}$ defined by
$(u,v)$-coordinates can be either compact or noncompact.
In flat ${\rm R}^{2}$ case ($f(u)=u^2$) tachyon tube solutions were
obtained in Ref.~\cite{Kim:2003uc}. Here
we will show that there exist various tachyon tubes with
arbitrary cross sectional shapes, as in the case of
supertubes~\cite{Bak:2001xx}. In particular, configurations on ${\rm S}^{2}$
will be considered in greater detail in the subsequent section.

As an ansatz, we assume that the fields are dependent only on the coordinate
$u$ and $F_{0u} = F_{uv} = F_{zu} = 0$.
Then nonvanishing fields are $T=T(u)$, $F_{0z}\equiv E_z(u)$,
$F_{0v}\equiv E_v(u)$, and $F_{v z}/f \equiv B_u (u)$.
With the ansatz, the Bianchi identity dictates $E_z$ and $E_v$ to be constants,
and $B_u\sim 1/f$. In this paper we further restrict our interest to looking
for the configurations with the critical value for $E_z$ and vanishing
$E_v$\footnote{Nonvanishing $E_{v}$ ($E_{v}^{2}<\alpha^{2}$)
can easily be understood through a boost transformation along $z$-direction
and the corresponding object is a helical tachyon tube~\cite{Kim:2003uc}.}
so that we have
\begin{equation}\label{Bian-2}
|E_z|=1,\qquad E_v = 0,\qquad B_u
= \frac{\alpha}{f},
\end{equation}
where $\alpha$ is an arbitrary D0 charge density
at $f=0$ and due to that the Bianchi identity $\nabla\cdot {\bf B}=0$
fails at $f=0$.

Substituting Eq.~(\ref{Bian-2}) with $T=T(u)$ into the action
(\ref{fa}), we find that the action is independent of the metric function
$f(u)$,
\begin{eqnarray}\label{rfa}
S= - {\cal T}_{3}\alpha \int dtdz dv \int du\, V(T)\sqrt{1+T'^{2}},
\end{eqnarray}
where the prime denotes differentiation with respect to
the variable $u$. Then the equation of motion reduces to
\begin{equation}\label{trr}
-fT_{uu} \equiv
{\cal T}_{3}\frac{V\alpha}{\sqrt{1+T'^{2}}} = \beta\alpha^{2}\, ,
\end{equation}
where $\beta$ is a nonnegative constant and $T_{uu}$
the $uu$-component of pressure.

For the solutions of Eq.~(\ref{trr}), many components of energy-momentum
tensor $T_{\mu\nu}$ and conjugate momenta of the gauge field
$\Pi_{i}$ vanish,
\begin{equation}\label{zet}
T_{0z}=T_{0u}=T_{zu}=T_{zv}=T_{uv}
=T_{vv}=\Pi_{u}=\Pi_{v}=0.
\end{equation}
The nonvanishing components share the same functional $T$-dependence
(and hence the same $u$-dependence) except $T_{uu}$ in Eq.~(\ref{trr}),
\begin{eqnarray} \label{loq}
fT_{00} &=& (f^2 + \alpha^2) \Sigma(u), \nonumber \\
T_{0v} &=& \alpha f \Sigma(u),\nonumber  \\
T_{zz} &=& -f\Sigma(u), \nonumber \\
\Pi &=& f^2\Sigma(u),
\end{eqnarray}
where $\Pi \equiv \Pi_{z}$ and
\begin{equation} \label{Sigma}
\Sigma(u) = \beta (1+T'^{2})=\frac{1}{\beta\alpha^{2}}({\cal T}_{3}V)^{2}.
\end{equation}
The energy per unit length then satisfies the relation
\begin{equation} \label{BPS}
{\cal E} =\int du dv \, fT_{00} = Q_{{\rm F1}} + \int du dv \, \Pi B^{2}, 
\end{equation}
where $Q_{\rm F1}$ is F1 charge per unit length,
\begin{equation}
Q_{{\rm F1}} = \int du dv\, \Pi.
\end{equation}
Note that Eq.~(\ref{BPS}) holds irrespective of the form of both the tachyon
potential $V(T)$ and the metric function $f(u)$. We will shortly see that, 
with the 1/cosh-type potential (\ref{V3}) (and only with this potential),
the second term of Eq.~(\ref{BPS}) is identified as D0 charge per unit length.

Now let us discuss the solution of Eq.~(\ref{trr}) in detail. With the form 
of tachyon potential (\ref{V3}), it is easy to obtain the exact solution 
\begin{equation}\label{tub}
\sinh\left(\frac{T(u)}{R}\right)
=\pm \left[\sqrt{\left(
\frac{{\cal T}_{3}}{\alpha\beta}\right)^{2}-1}
\,\cos \left(\frac{u}{R}\right)\right],
\end{equation}
where we imposed the condition $T'(0)=0$ for regularity.
This solution represents a coaxial array of tubular kink-antikink
with periodicity $2\pi R$. Note that the period
is independent of integration constants
$\alpha$ and $\beta$. This is consistent with the discussion on the unique
property of 1/cosh tachyon potential in Sec.~2.

For the solution, the quantity $\Sigma(u)$ of Eq.~(\ref{Sigma}) is given by
\begin{equation}\label{loq2}
\Sigma(u) = \beta \frac{({\cal T}_3/\alpha\beta)^2}%
        {1 +\left[({\cal T}_3/\alpha\beta)^2 -1 \right] \cos^2 (u/R)}.
\end{equation}
The energy (tube tension) of a single kink (per unit length) is then
calculated as
\begin{eqnarray}\label{lene}
{\cal E}_{2}^{(n)}&=&\int dv\int_{(n-1)\pi R}^{n\pi R}du\, fT_{00} \nonumber\\
&=&\beta \int dv\int_{(n-1)\pi R}^{n\pi R} \ du \
  \frac{({\cal T}_3/\alpha\beta)^2 (f^2 + \alpha^2)}{
 1 +\left[({\cal T}_3/\alpha\beta)^2 -1 \right] \cos^2 (u/R)},
\end{eqnarray}
and the string charge per unit length is
\begin{eqnarray}\label{lsc}
Q_{{\rm F1}}^{(n)}&=&\int dv\int_{(n-1)\pi R}^{n\pi R}du \,\Pi
\nonumber\\
&=&\beta \int dv\int_{(n-1)\pi R}^{n\pi R} \ du \
  \frac{({\cal T}_3/\alpha\beta)^2 \ f^2 }{
 1 +\left[({\cal T}_3/\alpha\beta)^2 -1 \right] \cos^2 (u/R)}.
\end{eqnarray}
Though the energy and the string charge are not calculable explicitly
for general $f(u)$, the difference ${\cal E}_2^{(n)} - Q_{\rm F1}^{(n)}$ 
is quite simple and can be calculated explicitly,
\begin{eqnarray}\label{qd}
{\cal E}_2^{(n)} - Q_{\rm F1}^{(n)} &=& 
 \beta\alpha^2 \int dv\int_{(n-1)\pi R}^{n\pi R} \ du \
  \frac{({\cal T}_3/\alpha\beta)^2 }{1 +\left[({\cal T}_3 /
  \alpha\beta)^2 -1 \right] \cos^2 (u/R)}
\nonumber\\
&=&\pi \alpha R {\cal T}_3\int dv \nonumber \\
&\equiv& Q_{{\rm D0}}^{(n)} ,
\end{eqnarray}
which coincides with the D0-brane charge per unit length. Note that it
is independent of $f(u)$ or $\beta$. Therefore we have a BPS-like
sum rule
\begin{equation}\label{ld0}
{\cal E}_{2}^{(n)}= Q_{{\rm F1}}^{(n)}+Q_{{\rm D0}}^{(n)}.
\end{equation}
In addition, each unit tube (or antitube) carries angular momentum
per unit length
\begin{equation}\label{ang}
L^{(n)}
=-\alpha\beta\int dv\int_{(n-1)\pi R}^{n\pi R}du
\frac{({\cal T}_3/\alpha\beta)^2 \ f^2 }{
 1 +\left[({\cal T}_3/\alpha\beta)^2 -1 \right] \cos^2 (u/R)},
\end{equation}
which is proportional to the string charge, i.e.,
$L^{(n)}=-\alpha Q_{{\rm F1}}^{(n)}$.

It is well-known that the supertube solution of cylindrical symmetry
is a BPS object preserving 1/4-supersymmetry~\cite{Mateos:2001qs}
and this BPS nature is not disturbed for tubular branes with
arbitrary cross sectional shape~\cite{Bak:2001xx}.
In the above, we obtained the tachyon tube solution
for which the $u$-coordinate dependence
is arbitrary.
Since it is given by
the configuration of coaxial array of tube-antitubes with nonzero thickness
(\ref{tub}), it may not be a BPS object despite of the BPS like sum
rule (\ref{ld0}).
To see whether the configuration is a BPS object, we look into
the stress components on $(u,v)$-plane. From Eqs.~(\ref{zet}) and (\ref{trr}),
we find that $T_{uv}$ and $T_{vv}$ vanish but $T_{uu}$ does not.
If we accept vanishing of all stress components on
$(u,v)$-plane
as a strict saturation of the BPS bound of these spinning tachyon tubes,
it can be achieved in the limit either
$\alpha\rightarrow 0$ or $\beta\rightarrow 0$.
The former is a trivial limit of a fundamental string without D0's and is
of no interest, while the latter corresponds to the zero thickness limit of
the tachyon tube which becomes the supertube for ${\cal M}=$S${}^{2}$.
Among the other components which are not in the $(u,v)$-plane, $T_{zu}$
and $T_{zv}$ vanish before taking the zero-thickness BPS limit as in 
Eq.~(\ref{zet}). On the other hand, the nonvanishing ones in Eq.~(\ref{loq})
become delta functions since 
\begin{equation}
\Sigma(u) \stackrel{\beta\rightarrow0}{\longrightarrow}
\frac{\pi R{\cal T}_3}{\alpha}
\sum_n \delta\left(u-\left(n-\frac12\right)\pi R \right).
\end{equation}

\setcounter{equation}{0}
\section{Tachyon Tubes in the Background of Salam-Sezgin Vacuum}

Here we study the case ${\cal M}_2 = $S${}^{2}$ in more detail.
Spheres appear to be a possible candidate for internal space
and well-known examples involving S${}^{2}$ include
AdS${}_{2}\times$S${}^{2}$ and compactifications on some Calabi-Yau
manifolds with S${}^{2}$ as a submanifold.
For simplicity, we assume that the other directions are flat, so the
background geometry of our interest is R${}^{1,1}\times$S${}^{2}$.
A representative example relevant with this flat space
is $N = 2$ Einstein-Maxwell
supergravity in six-dimensional space ${\rm R}^{1,3} \times
{\rm S}^2$ which is known as Salam-Sezgin model~\cite{Salam:1984cj}.
Its low energy limit
admits four-dimensional $N=1$ supergravity
which includes chiral fermions and of which the gauge symmetry is
SO(3)$\times$U(1). The geometry R${}^{1,3}\times$S${}^{2}$
of the Salam-Sezgin vacuum is expressed by
\begin{equation}\label{ssbg}
ds_{6}^{2}=-dt^{2}+dx^{2}+dy^{2}+dz^{2}+\frac{1}{8g^2}(d\theta^{2}
+\sin^{2}\theta\, d\varphi^{2}),
\end{equation}
where $0\le \theta \le \pi$ and
$0\le\varphi\le 2\pi$. 
There is a constant magnetic field $-1/2g$ on the two sphere inversely 
proportional to the gauge coupling $g$ of the Salam-Sezgin model.
This $N=2$ supergravity on R${}^{1,3}\times$S${}^{2}$ and its
variants have recently
attracted attention in relation with various
topics~\cite{Aghababaie:2002be,Aghababaie:2003wz,Guven:2003uw,Gibbons:2003gp,
Kerimo:2003am,Gibbons:2003di,Aghababaie:2003ar,Ghoshal:2003jd,
Kerimo:2004md,Cvetic:2003xr}.

Motivated by the above, we
consider a tachyon tube-antitube solution (\ref{tub}) on
R${}^{1,1}\times$S${}^{2}$ described by the coordinates 
$(t,z,\theta ,\varphi )$
embedded in the Salam-Sezgin vacuum, R${}^{1,3}\times$S${}^{2}$ (\ref{ssbg}).
Since the period of the solution is $2\pi R$ from Eq.~(\ref{tub}),
we identify the coordinates in the background metric (\ref{met}) as
\begin{equation}\label{resc}
u= R\theta,\quad v=R\varphi,\quad f=\sin\theta=\sin \left(
\frac{u}{R}\right),
\end{equation}
with $0\le u \le \pi R$ and $0\le v \le 2\pi R$. 
Then, $g$ is identified as 
$g=1/(2\sqrt{2}R)$, and 
the resultant background metric becomes Eq.~(\ref{met}).
If the radius $R$ introduced through the tachyon potential (\ref{V3}) has 
a string origin like $R=\sqrt{2}$ or $R=2$, the gauge coupling $g$ is of
the string scale.

\begin{figure}
\centerline{\epsfig{figure=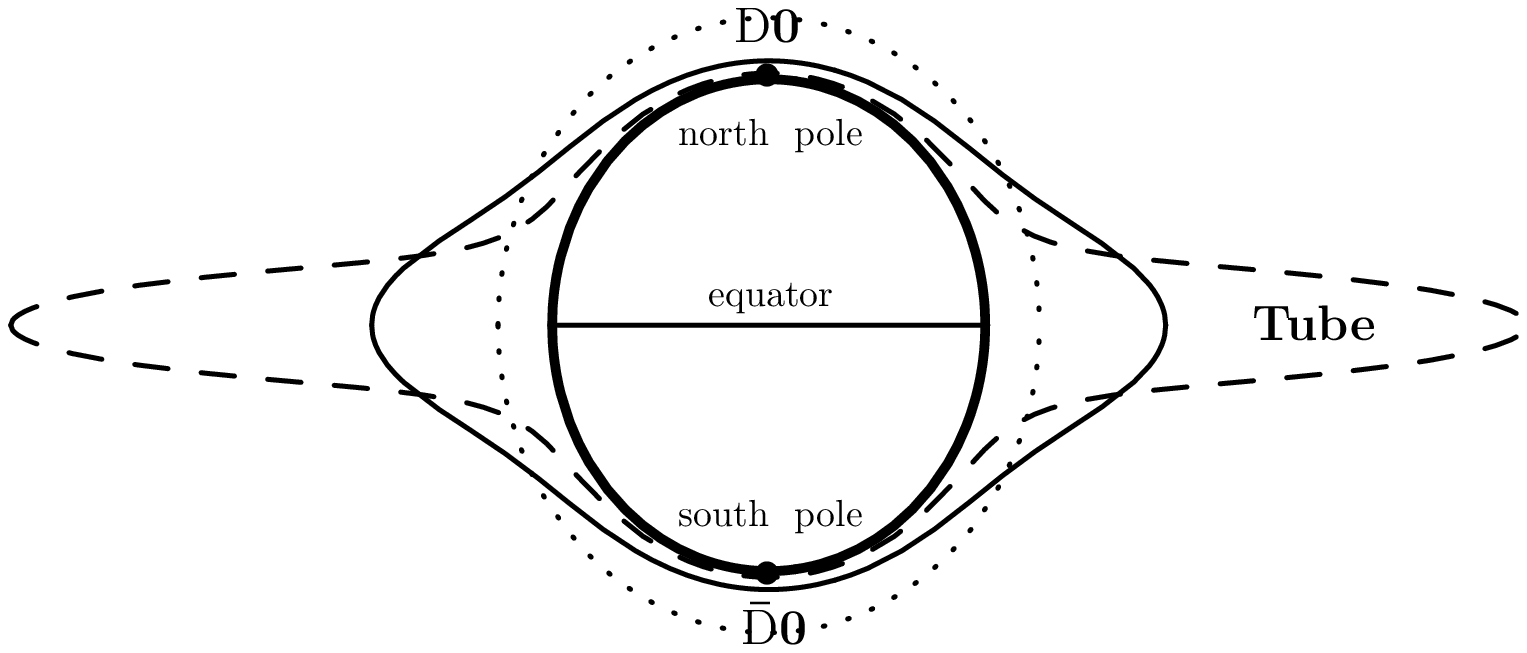,height=50mm}}
\caption{Plots of $\Pi (u)$ and ${\cal E}(u) -\Pi(u)$ for $\alpha
=1$ on S${}^{2}$: dashed line for $\alpha\beta/{\cal T}_{3}=0.1$,
solid line for $\alpha\beta/{\cal T}_{3}=0.3$, and dotted line for
$\alpha\beta/{\cal T}_{3}=1$. The profiles along the equator
represent the tubes and two peaks on both the north and south
poles D0 and $\bar{{\rm D}}0$.} \label{fig1}
\end{figure}

{}From the obtained tachyon profile (\ref{tub}),
we read that a single tachyon tube lies along the equator $(u=\pi R/2)$
and thereby F1 charge density is
accumulated there (See Figure~\ref{fig1}).
Linear D0's along $z$-axis are located at the north pole $(u=0)$ and
$\bar{{\rm D}}0$'s at the south pole $(u=\pi R)$.
An intriguing point is that the energy and the F1 charge per unit length
are obtained in closed forms
\begin{eqnarray}
{\cal E}_{2}&=&\beta \int_{0}^{2\pi R} \ dv \
\int_{0}^{\pi R} \ du \
  \frac{({\cal T}_3/\alpha\beta)^2 \left[\sin^2 (u/R) + \alpha^2\right]}{
 1 +\left[({\cal T}_3/\alpha\beta)^2 -1 \right] \cos^2 (u/R)}
 \nonumber\\
&=&2\pi R\left[\alpha + \frac{1}{\alpha}\frac{1}{1+(\alpha\beta/{\cal T}_3) }
\right]\times \pi R{\cal T}_{3},
\label{Slene}\\
Q_{{\rm F1}}&=&
\beta \int_{0}^{2\pi R} \ dv \
\int_{0}^{\pi R} \ du \
  \frac{({\cal T}_3/\alpha\beta)^2 \sin^2 (u/R)}{
 1 +\left[({\cal T}_3/\alpha\beta)^2 -1 \right] \cos^2 (u/R)}
\nonumber\\
&=& \frac{2\pi R}{\alpha}\frac{1}{1+(\alpha\beta/{\cal T}_3) }
\times \pi R{\cal T}_{3}.
\label{Slsc}
\end{eqnarray}
In the thin limit $(\alpha\beta/{\cal T}_3\rightarrow 0)$
of a single tachyon tube on S${}^{2}$ of radius $R$, F1 charge density is
concentrated along the equator
like the ring of the Saturn (see the solid and dashed lines
in Figure~\ref{fig1}). In the opposite limit
$(\alpha\beta/{\cal T}_3\rightarrow 1)$ with $T(u)=0$ at everywhere,
${\cal E}(u) -\Pi(u)$ is evenly distributed (see the dotted line
in Figure~\ref{fig1}).
Locations of two point-like peaks due to D0 and $\bar{{\rm D}}0$ are
also indicated at
both the north and the south poles, respectively in Figure~\ref{fig1}.

Product of two-dimensional flat directions $R^{2}$ to
${\rm R}^{1,1}\times {\rm S}^2$ is automatic so that the obtained
tube solution on ${\rm R}^{1,1}\times
{\rm S}^{2}$ is a tachyon tube solution in Salam-Sezgin model of
${\rm R}^{1,3}\times {\rm S}^{2}$. Therefore, it describes either a formation
of tubular D2-brane from an unstable D3-brane wrapped on ${\rm R}^{1}\times
{\rm S}^{2}$ or that of tubular D4-brane from the space-filling
unstable D5-brane in Salam-Sezgin model.

\setcounter{equation}{0}
\section{Conclusion}

In this paper we studied DBI-type effective theory of unstable D3-branes
and obtained an exact tubular D2-brane solution of arbitrary cross sectional
shape. The background manifold of the solution is 
${\rm R}^{1,1}\times {\cal M}_2$ where ${\cal M}_2$ is an arbitrary
two-dimensional manifold.
As the tachyon potential, we employed $1/\cosh$ potential. It
was shown that it has a unique property that an array of tachyon soliton 
solutions has a fixed period which is independent of integration constants of
the equations of motion. It also allows us to obtain closed form of solutions.
The thin BPS limit of the configurations
leads to supertubes of arbitrary cross sectional shapes.
In particular, we investigated the case ${\cal M}_2={\rm S}^{2}$ 
in more detail for which
the solution is embedded in the background geometry ${\rm R^{1,3}} \times 
{\rm S}^2$
of Salam-Sezgin model. Since a lifting of this model to ten-dimensional
type I supergravity is made, of which weak string coupling limit coincides with
an exact string theory solution, the near-horizon geometry of a
Neveu-Schwarz (NS) five-brane~\cite{Cvetic:2003xr}, it would be intriguing
to find 9-dimensional analogue of our solution 
in this 10-dimensional background.

Though our discussions were only about static objects, dynamical 
generation of D$(p-1)\bar{{\rm D}}(p-1)$ or tubular solution
should be achieved as inhomogeneous time-dependent solutions~\cite{Sen:2003zf}.
Until now, it seems incomplete 
since the solution seems to hit a singularity after time evolution for a finite 
time~\cite{Sen:2002vv}.

\section*{Acknowledgements}
The authors are indebted to Ashoke Sen for his valuable
suggestions and comments.
We also would like to thank Jin-Ho Cho, Seongtag Kim, Sangmin Lee, 
Hyeonjoon Shin,
and J. Troost for helpful discussions.
This work was supported by grant No.\ R01-2003-000-10229-0 from
the Basic Research Program of the Korea Science $\&$ Engineering Foundation
(C.K.) and is the result of research activities (Astrophysical Research
Center for the Structure and Evolution of the Cosmos (ARCSEC))
supported by Korea Science $\&$ Engineering Foundation(Y.K. and O.K.).


\begin{thebibliography}{99}

\bibitem{Sen:1998ex}
A.~Sen,
``BPS D-branes on non-supersymmetric cycles,''
JHEP {\bf 9812}, 021 (1998)
[arXiv:hep-th/9812031].

\bibitem{Horava:1998jy}
P.~Horava,
``Type IIA D-branes, K-theory, and matrix theory,''
Adv.\ Theor.\ Math.\ Phys.\  {\bf 2}, 1373 (1999)
[arXiv:hep-th/9812135].

\bibitem{Sen:1998tt}
A.~Sen,
``SO(32) spinors of type I and other solitons on brane-antibrane pair,''
JHEP {\bf 9809}, 023 (1998)
[arXiv:hep-th/9808141].

\bibitem{Sen:1999md}
A.~Sen,
``Supersymmetric world-volume action for non-BPS D-branes,''
JHEP {\bf 9910}, 008 (1999)
[arXiv:hep-th/9909062].

\bibitem{Garousi:2000tr}
M.~R.~Garousi,
``Tachyon couplings on non-BPS D-branes and Dirac-Born-Infeld action,''
Nucl.\ Phys.\ B {\bf 584}, 284 (2000)
[arXiv:hep-th/0003122];

E.~A.~Bergshoeff, M.~de Roo, T.~C.~de Wit, E.~Eyras and S.~Panda,
``T-duality and actions for non-BPS D-branes,''
JHEP {\bf 0005}, 009 (2000)
[arXiv:hep-th/0003221];

J.~Kluson,
``Proposal for non-BPS D-brane action,''
Phys.\ Rev.\ D {\bf 62}, 126003 (2000)
[arXiv:hep-th/0004106].

\bibitem{Buchel:2002tj}
A.~Buchel, P.~Langfelder and J.~Walcher,
``Does the tachyon matter?,''
Annals Phys.\  {\bf 302}, 78 (2002)
[arXiv:hep-th/0207235].

\bibitem{Kim:2003he}
C.~Kim, H.~B.~Kim, Y.~Kim and O-K.~Kwon,
``Electromagnetic string fluid in rolling tachyon,''
JHEP {\bf 0303}, 008 (2003)
[arXiv:hep-th/0301076];

C.~Kim, H.~B.~Kim, Y.~Kim and O-K.~Kwon,
``Cosmology of rolling tachyon,''
proceedings of PAC Memorial
Symposium on Theoretical Physics, pp.209-239 (Chungbum Publishing House,
Seoul, 2003),
arXiv:hep-th/0301142.

\bibitem{Leblond:2003db}
F.~Leblond and A.~W.~Peet,
``SD-brane gravity fields and rolling tachyons,''
JHEP {\bf 0304}, 048 (2003)
[arXiv:hep-th/0303035].

\bibitem{Lambert:2003zr}
N.~Lambert, H.~Liu and J.~Maldacena,
``Closed strings from decaying D-branes,''
arXiv:hep-th/0303139.

\bibitem{Kim:2003in}
C.~Kim, Y.~Kim and C.~O.~Lee,
``Tachyon kinks,''
JHEP {\bf 0305}, 020 (2003)
[arXiv:hep-th/0304180].

\bibitem{Kutasov:2003er}
D.~Kutasov and V.~Niarchos,
``Tachyon effective actions in open string theory,''
Nucl.\ Phys.\ B {\bf 666}, 56 (2003)
[arXiv:hep-th/0304045].

\bibitem{Sen:2003bc}
A.~Sen,
``Open and closed strings from unstable D-branes,''
Phys.\ Rev.\ D {\bf 68}, 106003 (2003)
[arXiv:hep-th/0305011].

\bibitem{Kim:2003ma}
C.~Kim, Y.~Kim, O-K.~Kwon and C.~O.~Lee,
``Tachyon kinks on unstable Dp-branes,''
JHEP {\bf 0311}, 034 (2003)
[arXiv:hep-th/0305092].

\bibitem{Brax:2003rs}
P.~Brax, J.~Mourad and D.~A.~Steer,
``Tachyon kinks on non BPS D-branes,''
Phys.\ Lett.\ B {\bf 575}, 115 (2003)
[arXiv:hep-th/0304197].

\bibitem{Sen:2003tm}
A.~Sen,
``Dirac-Born-Infeld action on the tachyon kink and vortex,''
Phys.\ Rev.\ D {\bf 68}, 066008 (2003)
[arXiv:hep-th/0303057].

\bibitem{Mateos:2001qs}
D.~Mateos and P.~K.~Townsend,
``Supertubes,''
Phys.\ Rev.\ Lett.\  {\bf 87}, 011602 (2001)
[arXiv:hep-th/0103030].

\bibitem{Bak:2001xx}
D.~Bak and A.~Karch,
``Supersymmetric brane-antibrane configurations,''
Nucl.\ Phys.\ B {\bf 626}, 165 (2002)
[arXiv:hep-th/0110039].

\bibitem{Kim:2003uc}
C.~Kim, Y.~Kim, O-K.~Kwon and P.~Yi,
``Tachyon tube and supertube,''
JHEP {\bf 0309}, 042 (2003)
[arXiv:hep-th/0307184].

\bibitem{Salam:1984cj}
A.~Salam and E.~Sezgin,
``Chiral compactification on Minkowski X S**2 of N=2 Einstein-Maxwell
supergravity in six-dimensions,''
Phys.\ Lett.\ B {\bf 147}, 47 (1984).

\bibitem{Aghababaie:2002be}
Y.~Aghababaie, C.~P.~Burgess, S.~L.~Parameswaran and F.~Quevedo,
``SUSY breaking and moduli stabilization from fluxes in gauged 6D
supergravity,''
JHEP {\bf 0303}, 032 (2003)
[arXiv:hep-th/0212091].

\bibitem{Aghababaie:2003wz}
Y.~Aghababaie, C.~P.~Burgess, S.~L.~Parameswaran and F.~Quevedo,
``Towards a naturally small cosmological constant from branes in 6D
supergravity,''
Nucl.\ Phys.\ B {\bf 680}, 389 (2004)
[arXiv:hep-th/0304256].

\bibitem{Guven:2003uw}
R.~Guven, J.~T.~Liu, C.~N.~Pope and E.~Sezgin,
``Fine tuning and six-dimensional gauged N = (1,0) supergravity vacua,''
Class.\ Quant.\ Grav.\  {\bf 21}, 1001 (2004)
[arXiv:hep-th/0306201].

\bibitem{Gibbons:2003gp}
G.~W.~Gibbons and C.~N.~Pope,
``Consistent S**2 Pauli reduction of six-dimensional chiral gauged
Einstein-Maxwell supergravity,''
arXiv:hep-th/0307052.

\bibitem{Kerimo:2003am}
J.~Kerimo and H.~Lu,
``New D = 6, N = (1,1) gauged supergravity with supersymmetric Minkowski(4) x
S**2 vacuum,''
Phys.\ Lett.\ B {\bf 576}, 219 (2003)
[arXiv:hep-th/0307222].

\bibitem{Gibbons:2003di}
G.~W.~Gibbons, R.~Guven and C.~N.~Pope,
``3-branes and uniqueness of the Salam-Sezgin vacuum,''
arXiv:hep-th/0307238.

\bibitem{Aghababaie:2003ar}
Y.~Aghababaie, C.P. Burgess, J.M. Cline, H. Firouzjahi, S. Parameswaran,
F. Quevedo, G. Tasinato, I. Zavala,
``Warped brane worlds in six dimensional supergravity,''
JHEP {\bf 0309}, 037 (2003)
[arXiv:hep-th/0308064].

\bibitem{Ghoshal:2003jd}
D.~Ghoshal, D.~P.~Jatkar and M.~Kreuzer,
``NS fivebrane and tachyon condensation,''
arXiv:hep-th/0312245.

\bibitem{Kerimo:2004md}
J.~Kerimo, J.~T.~Liu, H.~Lu and C.~N.~Pope,
``Variant N=(1,1) Supergravity and (Minkowski)${}_4$ x S${}^2$ Vacua,''
arXiv:hep-th/0401001.

\bibitem{Cvetic:2003xr}
M.~Cvetic, G.~W.~Gibbons and C.~N.~Pope,
``A string and M-theory origin for the Salam-Sezgin model,''
Nucl.\ Phys.\ B {\bf 677}, 164 (2004)
[arXiv:hep-th/0308026].

\bibitem{Gerasimov:2000zp}
A.~A.~Gerasimov and S.~L.~Shatashvili,
``On exact tachyon potential in open string field theory,''
JHEP {\bf 0010}, 034 (2000)
[arXiv:hep-th/0009103];

D.~Kutasov, M.~Marino and G.~W.~Moore,
``Some exact results on tachyon condensation in string field theory,''
JHEP {\bf 0010}, 045 (2000)
[arXiv:hep-th/0009148];

D.~Kutasov, M.~Marino and G.~W.~Moore,
``Remarks on tachyon condensation in superstring field theory,''
arXiv:hep-th/0010108.

\bibitem{Sen:2002an}
A.~Sen,
``Field theory of tachyon matter,''
Mod.\ Phys.\ Lett.\ A {\bf 17}, 1797 (2002)
[arXiv:hep-th/0204143].

\bibitem{Okuyama:2003wm}
K.~Okuyama,
``Wess-Zumino term in tachyon effective action,''
JHEP {\bf 0305}, 005 (2003)
[arXiv:hep-th/0304108].

\bibitem{Niarchos:2004rw}
V.~Niarchos,
``Notes on tachyon effective actions and Veneziano amplitudes,''
arXiv:hep-th/0401066.

\bibitem{Sen:2003zf}
A.~Sen,
``Moduli space of unstable D-branes on a circle of critical radius,''
JHEP {\bf 0403}, 070 (2004)
[arXiv:hep-th/0312003].

\bibitem{LLt} L.D. Landau and E.M. Lifshitz, section 12 of {\it Mechanics},
3rd ed, (Pergamon press, 1976).

\bibitem{Sen:2002vv}
A.~Sen,
``Time evolution in open string theory,''
JHEP {\bf 0210}, 003 (2002)
[arXiv:hep-th/0207105];

G.~N.~Felder, L.~Kofman and A.~Starobinsky,
``Caustics in tachyon matter and other Born-Infeld scalars,''
JHEP {\bf 0209}, 026 (2002)
[arXiv:hep-th/0208019];

J.~M.~Cline and H.~Firouzjahi,
``Real-time D-brane condensation,''
Phys.\ Lett.\ B {\bf 564}, 255 (2003)
[arXiv:hep-th/0301101];

S.~J.~Rey and S.~Sugimoto,
``Rolling of modulated tachyon with gauge flux and emergent fundamental
string,''
Phys.\ Rev.\ D {\bf 68}, 026003 (2003)
[arXiv:hep-th/0303133].

O-K.~Kwon and P.~Yi,
``String fluid, tachyon matter, and domain walls,''
JHEP {\bf 0309}, 003 (2003)
[arXiv:hep-th/0305229].


\end{thebibliography}
\end{document}